\documentclass[12pt]{article}
\makeatletter
\@addtoreset{equation}{section}

\newcommand{\be}{\begin{equation}} \newcommand{\ee}{\end{equation}}
\newcommand{\bea}{\begin{eqnarray}} \newcommand{\eea}{\end{eqnarray}}

\newcommand{\N}{{\cal N}}
\newcommand{\A}{{\cal A}}
\newcommand{\M}{{\cal M}}
\newcommand{\F}{{\cal F}}

\newcommand{\psib}{\bar{\psi}}

\newcommand{\ra}{\rightarrow}
\newcommand{\wg}{\wedge}

\newcommand{\R}{{\cal R}}
\newcommand{\cN}{{\cal N}}
\newcommand{\cD}{{\cal D}}
\newcommand{\cM}{{\cal M}}

\newcommand{\Tr}{\mbox{Tr}}

\begin{document}

\begin{titlepage}
\hfill
\vbox{\halign{#\hfil          \cr
             TAUP/2698-02     \cr
             CERN-TH/2002-037 \cr
             hep-th/0203092   \cr}} 
\vspace*{20mm}
\begin{center}
{\Large {\bf  Branes in Special Holonomy Backgrounds}\\} 
\vspace*{15mm}
\vspace*{1mm}
Amit Loewy$^a$ and Yaron Oz$^{a,b}$ \\
\vspace*{1cm} 
{\it {$^{a}$ School of Physics and Astronomy,\\
Raymond and Beverly Sackler Faculty of Exact Sciences\\
Tel-Aviv University, Ramat-Aviv 69978, Israel}}\\ 

\vspace*{5mm}
{\it {$^{b}$Theory Division, CERN \\
CH-1211 Geneva  23, Switzerland}}\\

\vspace*{.75cm}
\end{center}

\begin{abstract}
We study the flow from the theory of D2-branes in a $G_2$ holonomy background
to M2-branes in a $Spin(7)$ holonomy background. 
We consider in detail the UV and
IR regimes, and the effect of topology change of the background on the
field theory. We conjecture a non-Abelian $\cN=1$ mirror
symmetry.

\end{abstract}
\vskip 4cm

March  2002
\end{titlepage}
\newpage

\section{Introduction}

The study of string/M-theory backgrounds with special holonomy groups 
is receiving much attention lately.
One application of special holonomy manifolds is in 
constructing gravity duals of field theories with reduced supersymmetry,
and the analysis of the field theories on brane probes in such backgrounds.

In this letter we will consider two-brane probes in $Spin(7)$ and $G_2$
holonomy backgrounds.
Specifically, we will concentrate on two special manifolds with
$Spin(7)$ holonomy that were constructed in \cite{CGLP}; the $A_8$ and
$B_8$ manifolds. These non-compact manifolds
have two distinct features. While they have different topologies, they
admit the same metric up to a sign of a parameter. Asymptotically,
this metric is the same in both cases, and is locally
conical. 

We will analyze the dynamics of a system of M2-branes in these
$Spin(7)$ holonomy manifolds. We will describe the flow 
from the theory of D2-branes on a $G_2$ holonomy background
to M2-branes in these $Spin(7)$ holonomy backgrounds. We will consider
in detail the UV and IR regimes, and the effect of topology change 
(from $A_8$ to $B_8$) on the field theory.

The letter is organized as follows.
In the next section we will briefly review the $A_8$ and $B_8$ backgrounds.
In section 3 we will consider M2-branes in these backgrounds. We will 
take the field theory limit, study the phase structure of the systems, and 
compute the two-point function of the stress-energy tensor.
In section 4 we will construct the UV field theory and conjecture
a non-Abelian $\cN=1$ mirror symmetry.
In section 5 we will discuss some aspects of the IR regime. 

\section{$Spin(7)$ holonomy manifolds: $A_8$ and $B_8$}

Consider the following general ansatz for an eight-dimensional metric 
\be \label{genans}
ds_8^2 = h^2(r)dr^2 + a^2(r) \sigma^2+ b^2(r)(D\mu_i)^2 + c^2(r)d\Omega_4^2 \ .
\ee
We denote by $(\psi,\theta,\phi)$ the Euler angles, and $\ d\Omega_4^2 \ $
is the metric on the unit 4-sphere. Then $\ \sigma \ $ is given by
\be
\sigma \equiv d\phi + \A_1 = d\phi + \cos \theta d \psi - \mu_i A_i \ ,
\ee
where
\be
\mu_1=\sin \theta \sin \psi, \quad \mu_2=\sin \theta \cos \psi, 
\quad \mu_3=\cos \theta \ ,
\ee
and  $A_i$ is the gauge connection of an $SU(2)$ instanton on $S^4$. 
We define the covariant derivative, and instanton field strength as
\be
D\mu_i = d\mu_i +\epsilon_{ijk}A_j \mu_k \ , \qquad F_i = d A_i + 
\epsilon_{ijk} \ A^j \wedge A^k \ .
\ee

In the following we will consider a special case of the general metric 
ansatz that has $Spin(7)$ holonomy \cite{CGLP} (which is a special
case of a family of $Spin(7)$ metrics \cite{CGLP,CGLP2}) 
\be \label{newspin7}
ds_8^2 = \frac{(r+l)^2 dr^2}{(r+3l)(r-l)} + 
\frac{l^2(r+3l)(r-l) \sigma^2}{(r+l)^2} + \frac{1}{4}(r+3l)(r-l)(D\mu_i)^2 + 
\frac{1}{2}(r^2-l^2)d\Omega_4^2 \ .
\ee
$l$ is a real parameter whose significance will be discussed shortly.
When $\ l>0 \ $ the radial coordinate $r$ is in the range 
$\ l \le r < \infty \ $. 
This manifold is denoted as $A_8$.
When $\ l<0 \ $ the radial coordinate is in the range
$\ -3l \le r < \infty \ $. This manifold is denoted as $B_8$.

In the large $r$ region, both $A_8$ and $B_8$
have the same asymptotic form ${\cal M} \times S^1$,
where $S^1$ is parametrized by $\sigma$ with $g_{\sigma \sigma} 
\ra l^2$, and ${\cal M}$ is a cone over $CP^3$ with a $G_2$ 
holonomy metric \cite{GPP}
\be \label{g2}
ds_7^2 = dr^2 + r^2 \left(\frac{1}{4}(D\mu_i)^2 + \frac{1}{2}
d\Omega_4^2 \right) \ .
\ee
If we dimensionally reduce the asymptotic $\M \times S^1$ geometry 
along the $S^1$ direction, 
we get the seven-dimensional background (\ref{g2}) and a 2-form KK 
field strength 
\be
\F_2 \equiv d \A_1 = \frac{1}{2} \epsilon_{ijk} \mu^k D\mu^i 
\wedge D\mu^j - \mu_i F_i \ .
\ee
The dilaton is asymptotically constant.
The geometric symmetries of this background were analyzed in \cite{AW}.
The base of the cone $\M$ is the coset space 
\be 
{Sp(2) \over SU(2) \times U(1)} \ .
\ee 
Therefore, it is invariant under $Sp(2)$ 
action from the left, and $SU(2) \times U(1) \times Z_2$ from the right. The
$SU(2) \times U(1)$ acts trivially, and $Z_2$ acts as $\mu_i \ra 
-\mu_i$. 
These symmetries extend to the cone $\ \M \ $. These 
are not the symmetries of $\ \F_2 \ $.
It is clear that $\ \F_2 \ $ is not invariant 
under the $Z_2$ action. An $SU(2)$ instanton solution on $S^4$ 
has an 
$SU(2) \times U(1)$ symmetry group, and there is an additional 
$SO(3) \simeq SU(2)$ symmetry
from acting on the $i$ index. Therefore, the type IIA background is only
invariant under $SU(2) \times SU(2) \times U(1)$ transformations. 
    
The first correction to the
asymptotic geometry in the $l/r$ expansion is linear in $l$
and distinguishes between the two manifolds
\be
\delta ds^2 = \pm \frac{1}{2}|l|r (D\mu_i)^2 \ , 
\label{dg}
\ee   
where the $+$ sign is for $A_8$, and the $-$ sign for $B_8$. The correction 
(\ref{dg}) is non-normalizable, 
$||\delta g|| \ra \infty$, which implies that $l$ is a parameter in the dual 
field theory. 
The geometry of the two manifolds in the small $r$ region is quite
different. Near $r=l$ the $A_8$ metric reads 
\be
ds_8^2=d\rho^2+\frac{1}{4}\rho^2\left(\sigma^2 +(D\mu_i)^2 +
  d\Omega_4^2\right)\equiv
d\rho^2 + \rho^2 d \Omega_7^2 \ ,
\ee
where $\rho^2=4l(r-l)$ and $\ d \Omega_7^2 \ $ is the metric on the unit 
7-sphere.
Thus, $A_8$ is topologically  $R^8$, and near $\rho=0$ also geometrically. 
Near $r=-3l$ the $B_8$ metric reads 
\be
ds_8^2=d\rho^2 + \rho^2 d \Omega_3^2 + 4l^2 d \Omega_4^2 \ ,
\ee
where $\rho^2=-4l(r+3l)$, and $\ d \Omega_3^2 \ $ and 
$\ d \Omega_4^2 \ $ are the metrics 
on the unit 3-sphere and 4-sphere respectively. Thus, the $B_8$
manifold is topologically an $\ R^4 \ $ bundle over $\ S^4 \ $.
We see that as we vary the parameter $l$ from positive to negative
values we change the topology of the manifold.

\section{M2-branes in $A_8$ and $B_8$}  

In the following we will consider M2-branes in the $A_8$ 
and $B_8$ backgrounds.
We will first analyze the phase diagram 
of the system, as done for D2/M2-branes in flat space in \cite{IMSY}.  
The background of $N$ M2-branes takes the usual form \cite{CGLP}
\be \label{m2ona8}
ds^2=H^{-2/3} dx_{012}^2 + H^{1/3} ds_8^2 , 
\qquad C_3 = dx_0 \wedge dx_1 \wedge dx_2 H^{-1} \ ,
\ee
where $H(r)$ obeys the Laplace equation on $A_8$. In the $B_8$ case, 
the asymptotics is the same, but near the origin there is a subtlety, which we 
will address later. $H(r)$ reads 
\be \label{H}
H = 1 + Q \int_{r}^{\infty} \frac{dx}{(x-l)^4(x+3l)^2} \ ,
\ee
where $Q \sim l_p^6 N/l \ $.

We consider the low-energy limit $l_p \rightarrow 0 \ $, with $r$ and $l$
also taken to zero such that
\be
U \equiv \frac{lr}{l_p^3}, \qquad L N \equiv  \frac{l^2}{l_p^3}N \ , 
\label{low}
\ee
are kept fixed.
The ten-dimensional string coupling is 
\be 
g_s = \left(\frac{l}{l_p}\right)^{2/3} = \frac{l}{l_s} \ ,
\ee 
and the three-dimensional gauge coupling is $g_{YM}^2 \sim g_sl_s^{-1} \ $.
In terms of these parameters, (\ref{low}) reads 
\be
U = \frac{r}{l_s^2}, \qquad L N = \frac{l}{l_s^2}N \ , 
\ee
which are fixed as $l_s \rightarrow 0 \ $.
Note that $L \sim g_{YM}^2 \ $.

Consider the phase diagrams for two-branes in the $A_8$ and $B_8$ backgrounds.
On the supergravity (string) side there are two dimensionless 
expansion parameters: the effective
string coupling $e^{\phi}$ and  the curvature in string units 
$l_s^2{\cal R} \ $.
On the field theory side, the expansion in string coupling and the
curvature expansion correspond to the
$1/N$ expansion and strong coupling expansion in
$(g_{YM}^2NE^{-1})^{-1/2}$ 
respectively, with $E$ being an energy scale.
On the supergravity side there is an additional dimensionless
parameter $L/U \ $. 
On the field theory side it corresponds to $g_{YM}^2E^{-1} \ $.
 
\subsection{The $A_8$ case}
For large $\ U \ $, it can be seen from (\ref{H}) that 
the radius of the M-theory circle parametrized by $\sigma$ vanishes, and 
we need to dimensionally reduce to the type IIA description. 
The asymptotic form of $H$ is given by
\be
H = \frac{g_{YM}^2N}{l_s^4U^5} \left(\frac{1}{5} -\frac{1}{3}\frac{L}{U}
+\frac{13}{7}\frac{L^2}{U^2} + O\left(\frac{L^3}{U^3}\right)\right) \ ,
\ee
The asymptotic (large $U$) background is 
\bea \label{d2}
ds^2 &=& l_s^2 \left(\frac{U^{5/2}}{\sqrt{g_{YM}^2N}} dx_{012}^2 + 
\frac{\sqrt{g_{YM}^2N}}{U^{5/2}} ds_7^2 \right), 
\qquad e^{2\phi} = \sqrt{\frac{g_{YM}^{10}N}{U^5}} \ , \nonumber\\
\F_2 &=& \frac{1}{2} \epsilon_{ijk} \mu^i D\mu^j \wedge D\mu^k - \mu^i
F_i \ , \qquad \F_4=dx_0 \wedge dx_1 \wedge dx_2 \wedge dH^{-1}.
\eea 
This background
is the near-horizon limit of $N$ D2-branes transverse to the $G_2$ holonomy
background (\ref{g2}), with a 2-form KK field strength. 

We define the dimensionless expansion parameter 
$g_{eff}^2 = g_{YM}^2 NU^{-1} \ $.
The curvature $l_s^2\R$ and the effective string coupling of the background 
(\ref{d2}) read
\bea
l_s^2 \R &\sim& \frac{1}{g_{eff}} = \sqrt{\frac{U}{g_{YM}^2N}} \ ,
\nonumber\\
 e^{\phi} &\sim& \frac{g_{eff}^{5/2}}{N} \ .
\label{Rp}
\eea 
In this notation $L/U \sim g_{eff}N^{-1} \ $.

To leading order in $L/U$ the phase diagram is the standard one \cite{IMSY}.
When $g_{eff} \ll 1$ we can use the perturbative field theory
description. This is the high energy regime
and the field theory has ${\cal N}=1$ supersymmetry. 
The transition to the type IIA supergravity description
is at $g_{eff} \sim 1 \ $. This occurs at $U \sim g_{YM}^2 N \ $. 
We can use the type IIA description as long as the effective string
coupling is small $e^{\phi}\ll 1 \ $.
The metric describes a system of $N$ D2-branes transverse to
a $G_2$ holonomy manifold.
When  $e^{\phi}\sim 1$ we have to use the eleven dimensional
description. This occurs at energy $U\sim g_{YM}^2 N^{1/5} \ $.
In the extreme IR the uplifted metric approaches the  $AdS_4 \times
S^7$ metric and we have the ${\cal N}=8$ SCFT description.

This picture is not precise when the corrections in $L/U$ are
taken into account. The leading correction to the asymptotic geometry
in $L/U$ is non-normalizable, 
$||\delta g|| \ra \infty \ $. This means that in the dual field theory
description $L$ should be understood as a parameter rather than as 
a vev of a field. This is indeed the case, $L \sim g_{YM}^2 \ $.
Near $U=L$ the harmonic function has the following expansion in $(U-L)/L$ 
\be
H = 
\frac{N}{l_p^3 U^3} 
\left(1 -\frac{3}{4}\frac{U}{L} + \frac{9}{16} \frac{U^2}{L^2} 
+O\left(\frac{U}{L}\right)^3 \right) \ ,
\ee 
where we redefined $U-L$ as $U \ $.
At $U=0$ the metric (\ref{m2ona8})  
becomes $AdS_4 \times S^7$, with
$R_{sphere}=2R_{AdS} \sim l_p N^{1/6} \ $.
This is the same background as the near horizon limit of M2-branes in
flat space with corrections in $U/L \ $. 
 Note, however, that the expansion in $U/L$ is not really useful since
it requires $U \ll L $ in order to be valid.

As we noted before,
changing the sign of $L$ ($l$ before taking the field theory limit)
implies a topology change in the supergravity background. 
On the field theory side it corresponds 
to $g_{YM}^2 \rightarrow -g_{YM}^2 \ $, which is a sign of a phase transition.
In particular the singularity of the supergravity background at 
$L\rightarrow 0$ corresponds to a free field theory 
$g_{YM}^2 \rightarrow 0 \ $.
Note in comparison that M-theory on $G_2$ holonomy background 
yields a theory with four supercharges where holomorphy can be used to argue
for smooth interpolation between different geometrical backgrounds \cite{AW}.
Here we consider theories with two supercharges where 
holomorphy cannot be used, and indeed
the process of moving from one background to the other is not smooth.

\subsection{The $B_8$ case}
Consider the phase diagram of M2-branes on the $B_8$.
The asymptotic form of $B_8$ is the same as that of $A_8$ to leading 
order in the $L/U$ expansion. Thus, we expect the supergravity 
approximation to hold up to $U\sim g_{YM}^2N$ as in the $A_8$ case. 
The IR is quite 
different. If we change the sign of $L$ in (\ref{H}) the point $U=3L$ is a 
singularity. This singularity is
of the ``good'' kind \cite{MN} in the sense that the time 
component of the metric, $g_{00} \ $, is decreasing as we approach the 
singular point. The harmonic function near $U=3L$ reads
\be\label{b8h}
H =  \frac{N}{l_p^3L^2U} \left(1 -\left(\frac{13}{12}-\ln4 + \ln 
\frac{L}{U} \right) \frac{U}{L}+ \cdots \right) \ ,
\ee
where we shifted $U-3L$ to $U \ $. 
From the type IIA point of view the dilaton is not monotonic. 
It peaks at some intermediate point 
and then decreases as we approach the singularity, where it vanishes. 
Therefore, near the singular point we should use the ten-dimensional 
description.  

The dimensional reduction of the $B_8$ manifold to ten dimensions describes 
a D6-brane wrapped on an $S^4$ coassociative cycle in the seven-dimensional 
$G_2$ holonomy manifold (\ref{g2}). 
Adding D2-branes implies that we should describe this regime using the D2-D6
system wrapping $S^4 \ $. 
Note that the type IIA background is that of D2-branes
smeared on the $S^4 \ $. This system does not have a decoupling limit. 
Technically, this is because the power of $U$ in $H$ is not the 
same as in the $A_8$ case.  
Following \cite{PS}, we should use the solution of localized D2-branes,
which does have a decoupling limit, in which we also hold 
fixed $N_2 / N_6 \ $.
For small $r$ the $B_8$ is geometrically $R^4 \times S^4 \ $. 
The volume of the $S^4$ is proportional to $N_2/N_6 \ $.
Therefore, we expect that for $N_2 \gg N_6$ we can treat the $B_8$ 
manifold as $R^8$ with good accuracy in this small $r$ region. 
This will re-produce the phase diagram  
of the flat D2-D6 system in \cite{PS}.
Specifically,  we will have a background of the form $AdS_4 \times X^7 \ $,  
where $X^7$ is the $Z_{N_6}$ orbifold of $B_8 \ $. 
We can also reduce this background to ten dimensions, which will result in 
a fibered $AdS$ background.
In section 5 we will outline some features of the
field theory that governs the IR of this D2-D6 system.
 
\subsection{Two-point function}
In the following we will use the supergravity description in order
to compute the two-point function of the stress-energy tensor. 
We will be interested, in particular, in 
the $L$ dependent corrections.
The computation requires the solution of the graviton
equation. However, 
for certain polarization
and momentum vector, the graviton equation reduces to the minimally
coupled scalar equation.
A similar computation in the supergravity background of D2-branes has been done
in \cite{GO}.
The minimally coupled scalar equation is 
\be\label{wave1}
\frac{1}{\sqrt{g}}\partial_{\mu}\left(\sqrt{g}g^{\mu \nu}
\partial_{\nu}\right)\phi(U,x)=0 \ .
\ee
We take 
$$
\phi(U,x) = f(U) \exp(ikx).
$$
and we impose the boundary condition $f(U_{\epsilon})=1$ at the UV cut-off 
$U_{\epsilon} \ $. 
The two-point function is given by
\be \label{fluxt}
\langle T(k)T(-k) \rangle = \frac{N^2}{\lambda^2} \F_T \equiv
\frac{N^2}{\lambda^2} f(U)\sqrt{g}g^{UU}\partial_Uf(U) 
\Big|_{U_0}^{U_{\epsilon}} \ ,
\ee
where $\lambda = g_{YM}^2N$ is the 't Hooft parameter, and $U_0=L$ 
for the $A_8$ manifold and $U_0=3L$ for the $B_8$
manifold.
It is convenient to change variables to $z=(\lambda k^2)^{1/3}U^{-1} \ $, and 
re-scale $L$ as $L \ra (\lambda k^2)^{1/3}L \ $. 
Since we will only need to
know the small $z$ behavior of $f(z)$ we can solve (\ref{wave1}) order by 
order. After expanding the coefficients in (\ref{wave1}) we get 
\be\label{wave2}
f''(z) + \frac{1}{z}\Big(-4+2(Lz)-22(Lz)^2+\cdots
%50L^3z^3+\ldots
%-166L^4z^4+482L^5z^5-1462L^6z^6+4370L^7z^7
\Big)f'(z)
+ z \left(\frac{1}{5}-\frac{(Lz)}{3}
+\frac{93(Lz)^2}{35}+\cdots
%-\frac{119L^3z^3}{15}+\frac{10474L^4z^4}{315}
\right)f(z)=0 \ .
\ee
The solution of (\ref{wave2}) after imposing the boundary condition is given by
\be\label{solution}
f(z)=Cz^5\left(1-\frac{5}{3}Lz+O(z^2)\right)
-\frac{4}{105} L^2 z^5 \ln(z)\left(1+O(z)\right)
+1+\frac{1}{30}  z^3+ O(z^4) \ ,
\ee
where $C$ is an integration constant. It is fixed by the condition that
at $U=U_0$ the solution is smooth, $f'(U_0)=0 \ $.

We use (\ref{fluxt}) to calculate the flux factor, and extract the 
cut-off independent piece
\be\label{fluxc}
\F_T= \frac{N^2}{\lambda^2} \left(5C\lambda^{5/3}k^{10/3} 
-\frac{8}{63} L^2 k^2 \lambda \ln(k) + 
\mbox{analytic in $k$} \right) \ .
\ee
In coordinate space the two-point function that follows from (\ref{fluxt})
reads 
\be\label{2pf}
\langle T(0)T(x) \rangle \sim \frac{N^2}{\lambda^2}\left(
\frac{\lambda^{5/3}}{|x|^{19/3}} + \frac{\lambda L^2}{|x|^5}\right) \ .
\ee
The first term in (\ref{2pf}) has been obtained in \cite{GO} for the D2-branes
in flat space. The second term is the $L/U$ correction. 

At the crossover regime between the supergravity description and the 
field theory
description 
$g_{YM}^2N|x| \simeq 1$ we get
\be
\langle T(0)T(x) \rangle \sim \frac{N^2}{|x|^{6}} \ ,
\ee
which matches to a field theory with $N^2$ degrees of freedom.

Note that in this computation we did not find an odd power of
$L$ dependence that will be different in the $A_8$ and $B_8$ cases. 
However, the computation is valid only in the UV regime.
As we noted and will also discuss later, the IR dynamics of M2-branes on 
the two manifolds is quite different.

\section{The UV field theory}

The field theory in the UV is a three-dimensional $\cN=1$
supersymmetric gauge theory.
Let us first briefly review some aspects of  
$\cN=1$ supersymmetry in three dimensions.
The supersymmetry algebra reads
\be
\{Q_{\alpha},\bar{Q}_{\beta}\} = 2 \gamma_{\alpha\beta}^{\mu}P_{\mu},~~~
\mu=0,1,2,~~\alpha=1,2  \ .
\ee
$Q_{\alpha}$ are two real spinor supercharges, and $\gamma^0=\sigma_2,
 \gamma^1=i\sigma_3, \gamma^2=i\sigma_1 \ $.
Superspace is described by three space-times coordinates, and by two 
real anti-commuting variables $\theta_{\alpha} \ $.
A scalar superfield $\Phi$ takes the form
\be
\Phi = \phi +\bar{\theta}\chi + \frac{1}{2}\theta\bar{\theta}F \ ,
\ee
where $\phi$ is a real scalar, $\chi_{\alpha}$ is a real spinor 
and $F$ is an auxiliary field.
A vector multiplet  can be written as a real spinor
superfield $V_{\alpha}$ (in the Wess-Zumino
gauge) 
\be
V_{\alpha} = i(\gamma^{\mu}A_{\mu}\theta)_{\alpha} +
\frac{1}{2}\bar{\theta}\theta\lambda_{\alpha}
\ .
\ee
The field strength tensor is 
\be
F_{\alpha} = \lambda_{\alpha} -i(\gamma^{\mu}B_{\mu}\theta)_{\alpha} -
\frac{1}{2}\bar{\theta}\theta(\gamma^{\mu}\partial_{\mu}\lambda)_{\alpha}
\ ,
\ee
where $B_{\mu}=\frac{1}{2}\varepsilon_{\mu\nu\lambda}F^{\nu\lambda} \ $.
The supersymmetric action is
\be
S = \int d^3 x d^2 \theta \left(-\frac{1}{2}F_{\alpha}\cdot
  F_{\alpha}
+\frac{1}{2}(\cD\Phi)\cdot(\cD\Phi) + W(\Phi) \right) \ ,
\label{S}
\ee
where $\cD_{\alpha} = D_{\alpha} -i e V_{\alpha}$ is the super-covariant 
derivative.
The action (\ref{S}) contains kinetic terms for the scalar, vector and
spinor fields, Yukawa couplings and a scalar potential
\be 
V =  \frac{1}{2} g^{ij} \partial_i W \partial_j W \ ,
\ee
where $g^{ij}$ is the kinetic term metric.

As we discussed in the previous section, at large $r$ the metrics of
$A_8$ and $B_8$ have the asymptotic form $\cM\times S^1 \ $, where $\cM$ 
is a cone over $CP^3 \ $.
Let us look for an $\cN=1$ supersymmetric 
gauge theory that has this space as its classical moduli space of
vacua. 
Consider a supersymmetric $U(1) \times U(1)$ gauge theory with an $\cN=1$
vector multiplet and four $\cN=2$ chiral (eight $\cN=1$ scalar)
superfields $A_i, B_i, i=1,2$ with $U(1) \times U(1)$ charges $(1,-1)$ for
$A_i$ and $(-1,1)$ for $B_i \ $. 
As reviewed above, there is no D-term for $\cN=1$
supersymmetry in three dimensions. 
Thus, the classical moduli space of vacua is given by the space parametrized
by the scalar components of $A_i, B_i$ modulo the $U(1) \times U(1)$ 
action. 

With the above 
charges, $A_i$ and $B_i$ are neutral under the diagonal 
$U(1)$. We define
\be 
w_1=a_1,\ \ w_2=a_2, \ \ w_3=b_1^*, \ \ w_4=b_2^* \ ,
\ee
where $a_i, b_i$ are the scalar components of $A_i, B_i \ $.
The classical moduli space of vacua is
\be
(w_1,w_2,w_3,w_4) \simeq \lambda (w_1,w_2,w_3,w_4),~~~~|\lambda|=1 \ ,
\ee
which is a cone over $CP^3$. 
In three dimensions the gauge field $A_{\mu}$ is dual to a compact real
scalar $q$, $dq = *dA \ $.
The scalar $q$ parameterizes a circle $S^1 \ $.
Thus, the classical moduli space of the theory is $\cM \times S^1 \ $.

A generalization to $N$ D2-branes 
requires that as we separate the branes we get $N$ copies of the above 
moduli space.
Consider a 
$U(N) \times U(N)$ $\cN=1$ vector multiplets 
with $A_i, B_i, i=1,2$, $\cN=2$ chiral superfields 
in the $(N,\bar{N}), (\bar{N},N)$ representations of the gauge groups,
respectively.
When $A_i, B_i$ are diagonal with distinct eigenvalues, 
the gauge group is broken 
to $U(1)^N \times U(1)^N \ $. This is the case where 
all the D2-branes are separated. In this case we will get 
$N$ copies of $\cM \times S^1 \ $.
However, we have to make the off-diagonal fields massive.
A superpotential that is compatible with the symmetries 
and gives mass to the off-diagonal
fields via the Higgs mechanism is 
\be
W \sim \varepsilon^{ij}\varepsilon^{kl}\Tr \left(A_i B_k A_j B_l \right) \ .
\ee
Thus, we propose this as the UV field theory for $N$ D2-branes 
in the background $\cM \ $.

We note that the field theory is very similar to the 
four-dimensional $\cN=1$ field 
theory on the world-volume of D3-branes placed at the 
tip of the conifold \cite{KW}.
Compactification of that theory on a circle would give 
$\cN=2$ three dimensional
gauge theory with four supercharges. 
The theory considered here has only two supercharges,  
since the vector superfields are $\cN=1$ multiplets and not $\cN=2 \ $. 
  
It has been proposed in \cite{GT}
that theory on the world-volume of one D2-brane in the 
background $\cM$ (theory $B$)
is mirror to another
$\cN=1$ three-dimensional gauge theory (theory $A$), whose Coulomb branch
is $\cM$. Theory $A$ is constructed by placing a D2-branes 
in the ten-dimensional
background obtained by the reduction of $\cM$ on $S^1 \in S^4 \ $.
The ten-dimensional background corresponds to D6-branes intersecting at angles.
The field theory of one D2-brane in the intersecting D6-branes background
is an $\cN=1$  $U(1)$ gauge theory with three $\cN=2$ chiral 
superfields $\Phi^i$ (from the 2-2 strings)
and two $\cN=2$ hypermultiplets (from the 2-6 strings).

One expects the generalization of theory $A$
to $N$ D2-branes to be an $\cN=1$
$U(N)$ gauge theory with 
three $\cN=2$ chiral superfields $\Phi^i$ in the adjoint 
and two $\cN=2$ hypermultiplets in the fundamental representation.
We also expect a superpotential of the form
\be
W \sim \varepsilon_{ijk}\Tr \left(\Phi^i\Phi^j\Phi^k\right) \ .
\ee
When $\Phi^i$ are diagonal with distinct eigenvalues the $N$ D2-branes
are separated and their positions are given by the
eigenvalues. The gauge group is broken to $U(1)^{N} \ $. 
The scalars together with the duals to the $N$ photons parameterize a
$7N$-dimensional space, made of $N$ copies of a seven-dimensional
space that admits a metric of $G_2$ structure.

It is tempting to conjecture a non-Abelian version of the 
$\cN=1$ mirror symmetry,
and suggest that the non-Abelian theories $A$ and $B$ are mirror in the sense
that the Coulomb branch of $A$ matches the Higgs branch of $B$.

\section{The IR regime}

In this section we discuss some aspects of the IR regime.

\subsection{The $A_8$ case}     
As we saw, in the $A_8$ case there is a single IR fixed point where 
supersymmetry is enhanced back to $\N=8 \ $. 
The irrelevant operators that deform the $S^7$ are dual to 
scalars in $AdS_4$. The question is how many irrelevant operators are
turned on, and what are their conformal 
dimensions and representation under $SO(8) \ $. Fortunately, some of the work 
was already carried out in \cite{AR}, where a more symmetric deformation 
of $S^7$ was analyzed. 

Our starting point will be the one dimensional Lagrangian, from which one 
can derive the equations that define the $A_8$ manifold \cite{CGLP}. 
\begin{eqnarray}
L & = & 2\alpha'^2 + 12\gamma'^2 + 4\alpha' \beta' + 8\beta' \gamma' + 
16 \alpha' \gamma' \nonumber \\
  & + & \frac{1}{2}b^2c^4(4a^6+2a^4b^2-24a^4c^2-4a^2c^2-4a^2c^4+b^2c^4),
\end{eqnarray} 
where $a, b, c$ are as in (\ref{genans}), $h(r)=1 \ $,
and $\alpha'=\partial_r \ln(a)$ and the same for $\beta$ and $\gamma \ $.
We shall now modify this Lagrangian to include the $N$ M2-branes. 
Note that the original Lagrangian was derived for an eight-dimensional 
space, and in order to describe the supergravity solution of the M2-branes 
we should first add three flat directions, and then multiply 
by the appropriate factors of $H(r) \ $. 
The potential term $V(a,b,c)$ must be added with the contribution of 
the 4-form field strength 
\be \label{lag1}
\Delta V = a^8b^4c^{12}Q^2. 
%\frac{1}{2}b^2c^4(4a^6+2a^4b^2-24a^4c^2-4a^2c^2-4a^2c^4+b^2c^4) + 
%a^8b^4c^{12}Q^2.
\ee
We can change to a more familiar set of variables \cite{AR}, which will make 
the geometric picture more clear. (For $a=b$ we should 
get the Lagrangian of \cite{Page}.)  
We define:   
\begin{eqnarray}
a & = & 2^{1/6} \exp(-\frac{3}{4}u-2v+\frac{1}{3}w) \ , \nonumber \\
b & = & 2^{1/6} \exp(-\frac{3}{4}u-2v-\frac{2}{3}w) \ , \nonumber \\
c & = & 2^{1/6} \exp(-\frac{3}{4}u+\frac{3}{2}v) \ ,
\end{eqnarray}
and denote the 
four-dimensional metric by $g_{ij}$.
With the new variables the new Lagrangian is 
\begin{eqnarray} \label{lag}
L & = & \sqrt{g} \ \Big( \R_g -\frac{63}{2}(\partial u)^2-21(\partial v)^2+
\frac{4}{3}(\partial w)^2 - 2Q^2e^{-21u} \nonumber \\
  & - & e^{-9u-10v}(8e^{2w/3}+4e^{-4w/3})-48e^{-9u-3v}+e^{-9u+4v}
(2e^{-8w/3}-8e^{-2w/3}) \ \Big).
\end{eqnarray}
The geometric meaning of each of the 
three functions in now clear. Let us 
for the moment set $w=0$ ($a=b$) identically (not just at $r=0$). 
In this case (\ref{lag1}) is the action that governs 
deformations of $S^7$ that 
preserve an $SO(5)\times SO(3)$ isometry \cite{AR}. 
In particular, the potential has 
two critical points, that correspond to Einstein manifolds. One is the round 
$S^7$ and the other is the squashed 
$S^7$. The round $S^7$ is the solution given by  
\be \label{rs7}
v_0=w_0=0, \qquad u_0=\frac{\ln(Q^2/9)}{12} \ ,
\ee
and corresponds to an IR fixed point.
The squashed $S^7$ solution, that will not be of interest here, 
has $v \neq 0 \ $, and corresponds to a UV fixed point.

In our case the symmetry group is smaller, since $w \neq 0$ ($a \neq b$) 
except at $r=0 \ $. We consider fluctuations of $u, v, w$ around the round 
$S^7$ solution. 

In eleven dimensional supergravity compactified on $AdS_4 \times S^7$ 
there are 3 scalar KK towers denoted by $S_1, S_2, S_3 \ $. 
Their $SO(8)$ representations and 4 dimensional masses are given by 
\cite{BCERS}
\begin{eqnarray}\label{scalars}
S_1 & (n+2,0,0,0) & m^2=(n-2)^2-9 \nonumber \\
S_2 & (n-2,0,0,0) & m^2=(n+7)^2-9 \nonumber \\
S_3 & (n-2,2,0,0) & m^2=(n+3)^2-9. 
\end{eqnarray}
There are also 2 pseudo-scalar KK towers denoted by $P_1$ and $P_2$
\begin{eqnarray}\label{pscalars}
P_1 & (n,0,2,0) & m^2=(n+1)^2-9 \nonumber \\
P_2 & (n-2,0,0,2) & m^2=(n+5)^2-9. 
\end{eqnarray} 
Following the calculations in \cite{AR}, the masses of the the 3 scalars 
in (\ref{lag}) are given by (in $AdS_4$ mass units)
\be 
M^2_{ij} = \frac{1}{2} \partial^2_{i,j} V(u,v,w) |_{u_0,v_0,w_0} \ ,
\ee
after rescaling the fields so that the kinetic terms are canonically 
normalized. Specifically we get that
\be
M^2_{uu}=M^2_{vv}=M^2_{ww}=+16 \ .
\ee
By the AdS/CFT correspondence these 3 scalars are dual to irrelevant operators
of dimensions $\Delta =4 \ $  \cite{AOY}. 
The first scalar $u$ is just the over-whole volume of $S^7 \ $, and 
is identified with $n=7$ of $S_1$ \ . 
The second scalar $v$ can be identified as $n=2 \ $, of $S_3$ transforming as 
the ${\bf 300}$ of $SO(8)$ \cite{AR}. The third scalar $w$ is actually a 
pseudo-scalar $n=4 \ $, of $P_1 \ $, and is responsible for 
deforming the $S^3$ fibers. It transforms as the ${\bf 5775}$ of $SO(8) \ $. 
With all three irrelevant perturbations turned on there is no second 
non-trivial fixed point, which means that in the UV the system should be 
described by perturbative field theory as is indeed the case.

\subsection{The $B_8$ case}

As discussed before, the $B_8$ manifold reduced to ten dimensions
describes the background of
a D6-brane wrapped on an $S^4$ supersymmetric 4-cycle 
($Vol(S^4) \sim l^4$) in
a $G_2$ holonomy manifold \cite{CGLP}.
The manifold is the bundle of anti-self-dual two-forms over $S^4 \ $. 
Note that to get $N_6$ D6-branes one 
should consider a $Z_{N_6}$ orbifold of the $S^1$ on which we reduce.

Let us consider the D-branes field theory when 
adding D2-branes in this background.
At energies $E \ll l^{-1}$ the effective theory on the D6-brane 
world-volume is a
$U(1)$ $\cN=1$ supersymmetric gauge theory with a Chern-Simons term \cite{GS}
\be 
L = \frac{1}{4g_6^2} \int d^3x (\F^2 + i \psib \Gamma\cdot D \psi)
    + \frac{i(k+\frac{1}{2})}{4\pi} \int (\A \wg d \A + \psib \psi) \ .
\ee       
The gauge coupling $g_6^2 \sim g_s l_s^3 l^{-4} \ $.
The Chern-Simons term arises from the WZ term, when taking into account
the  half integral $G_4$ flux required for the consistency of the 
compactification on $B_8 \ $ \cite{GS}.
Since the $S^4$ is rigid there are no massless scalars to parameterize its 
embedding in the seven-dimensional space.

Adding $N_2$ D2-branes to the system means an additional $\cN=1$ 
$U(N)$ vector multiplet with 
gauge coupling $g_2^{2} \sim g_s l_s^{-1} \ $. There is one
 $\cN=2$ hypermultiplet that parameterizes the motion of the 
D2-branes in $S^4 \ $,
and three scalar superfields that parameterize the three directions
normal to the $S^4 \ $.
We need to study the system at low energies $E \ll g_2^2,g_6^2 \ $.
At these energies the kinetic terms are irrelevant. 
Since a D2-brane in D6-branes wrapping $S^4$ can be viewed as an instanton 
on $S^4$, it is plausible to expect the the IR
theory will be a conformal field theory on the Higgs branch, 
which is the moduli space
of $N_2$-instantons of $U(N_6)$ on $S^4$ \footnote{
For a somewhat similar discussion in another context see 
\cite{Witten:1997yu}.}. This is consistent with the fact that 
the supergravity background for small 
$U$ has an $AdS_4$ factor, as noted in section 3.2.

\vskip 1cm

\section*{Acknowledgements}

We would like to thank O. Aharony, H. Ita, O. Kenneth and T. Sakai 
for valuable discussions. 
This research is supported by the US-Israel Binational Science
Foundation.

%\newpage

\end{document}